\documentstyle{llncs}

\def\compoundrel#1\over#2{\mathpalette\compoundreL{{#1}\over{#2}}}
\def\compoundreL#1#2{\compoundREL#1#2}
\def\compoundREL#1#2\over#3{\mathrel
     {\vcenter{\hbox{$\buildrel{#1#2}\over{#1#3}$}}}}

\begin{document}

\title{Steering and isotope effects in the dissociative adsorption 
       of H$_2$/Pd(100)}

\author{Axel Gross and Matthias Scheffler}
\institute{Fritz-Haber-Institut, Faradayweg 4-6, 
D-14195 Berlin-Dahlem, Germany} 

\date{}

\maketitle

\vspace{-.7cm}

\begin{abstract}

The interaction of hydrogen with many transition metal surfaces 
is characterized by a coexistence of activated 
with non-activated paths to adsorption  with a broad distribution 
of barrier heights. By performing six-dimensional quantum dynamical 
and classical molecular dynamics calculations  
using the same potential energy surface derived 
from {\it ab initio} calculations for the system H$_2$/Pd(100)
we show that these features of the potential energy surface
lead to strong steering effects in the dissociative adsorption dynamics.
The adsorption dynamics shows only a small isotope effect which is purely
due to the quantum nature of hydrogen.

\end{abstract}


\section{Introduction}

It is a long-term goal in surface science to understand
catalytic reactions occuring at surfaces. Obviously, the single steps
of these often rather complicated processes are more effectively
studied at simple systems. 
The dissociative adsorption of molecules on surfaces is one of the
fundamental reaction steps occuring in catalysis. This establishes
its technological relevance and importance.
In particular, the dissociative 
adsorption and associative desorption of hydrogen on metal surfaces 
has served as a benchmark system, both experimentally and theoretically
(see, e.g., Refs.~\cite{Ren94,Hol94,Dar95,Gro96SS} and references therein). 
Since the mass mismatch between hydrogen and a metal substrate 
is rather large, the crucial process in the dissociative adsorption
for these particular systems is the conversion of 
translational and internal energy of the hydrogen molecule into 
translational and vibrational energy of the adsorbed hydrogen atoms. 
If in addition no surface rearrangement occurs upon adsorption,
the substrate degrees of freedom can be neglected and the dissociation
dynamics can be described in terms of potential energy surfaces (PES)
which take only the molecular degrees of freedom into account.

The PES for the dissociative adsorption of a diatomic molecule 
neglecting the substrate degrees of freedom is still six-dimensional.
These PESs now become available by elaborate density-functional
calculations \cite{Ham94,Whi94,Wil95,Wil96,Wie96}.
However, in order to understand the reaction dynamics one has to
perform dynamical calculations on these potentials.
Because of its light mass hydrogen has to be described quantum
mechanically. Only recently it has become possible to
perform dynamical calculations of the dissociative adsorption
and associative desorption where {\em all} degrees of freedom of
the hydrogen molecule are treated quantum mechanically \cite{Gro95}. 
These calculations have established the importance of high-dimensional
effects in the reaction dynamics.

For example, molecular beam experiments of the 
dissociative adsorption of H$_2$ on various
transition metal surfaces like Pd(100) \cite{Ren89}, Pd(111) and Pd(110)
\cite{Res94}, W(111) \cite{Ber92}, W(100) \cite{Ber92,But94,Aln89},
W(100)--c(2$\times$2)Cu \cite{But95} and Pt(100) \cite{Dix94}  
revealed that the sticking probability in these systems
initially decreases with increasing kinetic energy of the beam.
High-dimensional quantum dynamical calculations have shown that
steering effects can cause such an initial decrease in the
sticking probability \cite{Gro95,Kay95}; it is not necessarily
due to a precursor mechanism, as was widely believed. 

So far we have studied, besides the steering effect \cite{Gro95}, the 
influence of the molecular rotation and orientation \cite{Gro95,Gro96} 
and vibration \cite{Gro96b} on the sticking probability of
H$_2$/Pd(100).
In this paper we present a comparison of quantum and classical
dynamics for the dissociation of hydrogen on Pd(100) and
investigate isotope effects. In the next section
the theoretical background will be introduced before the results
of the dynamical calculations will be discussed.

\section{Theoretical background}

The potential energy surface of H$_2$/Pd(100)
has been determined using density-functional
theory together with the generalized gradient approximation (GGA) \cite{Per92} 
and the full-potential linear augmented plane wave method \cite{Bla93,Koh95}. 
{\em Ab initio} total energies have been evaluated for more than 250 
configurations \cite{Wil95} and have been parametrized in a suitable 
form for the 
dynamical calculations \cite{Gro95}.

\begin{figure}[t]
\unitlength1.cm
\begin{minipage}{5.cm}
   \begin{picture}(6.0,7.5)
      \includegraphics{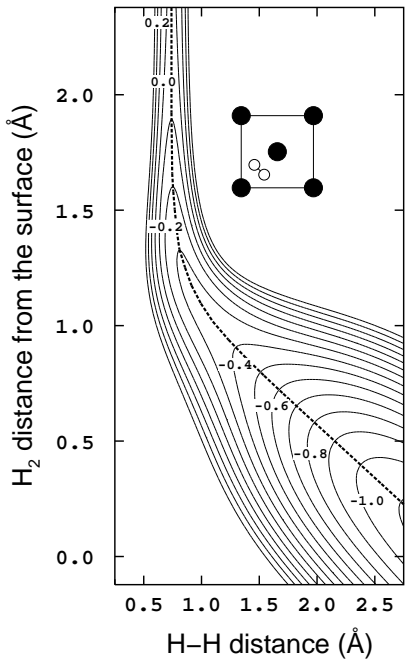}
   \end{picture}
\end{minipage}
\hspace{1.cm}
\begin{minipage}[b]{5.5cm}
   \caption{Contour plot of the PES along a
            two-dimensional cut through the
            six-dimensional coordinate space of H$_2$/Pd\,(100).
            The inset shows the
            orientation of the molecular axis and the lateral
            H$_2$ center-of-mass coordinates. The coordinates 
            in the figure are the H$_2$ center-of-mass distance 
            from the surface $Z$ and the H-H interatomic distance $d$. The 
            dashed line is the optimum reaction path.
            Energies are in eV per H$_2$ molecule.
            The contour spacing is 0.1~eV.  }
\label{elbow}
\end{minipage}
\end{figure}

Figure \ref{elbow} shows a cut through
the PES of H$_2$/Pd\,(100), where the most 
favourable path towards dissociative adsorption is marked by the dashed line. 
For this path there is no energy barrier hindering
dissociation, i.e., the adsorption is non-activated.
However, the majority of pathways towards dissociative adsorption
has in fact energy barriers with a rather broad
distribution of heights and positions, as the detailed total-energy 
calculations showed~\cite{Wil95}, i.e. the PES is strongly anisotropic 
and corrugated. This has important consequences, as will be shown below.

The quantum dynamics is determined in a coupled-channel
scheme within the concept of the {\em local reflection matrix} (LORE) 
\cite{Bre93}. This numerically very stable method
is closely related to the logarithmic derivative of the solution matrix
and thus avoids exponentially increasing outgoing waves which could
cause numerical instabilities. 
The reported calculations, which take all degrees of freedom of the hydrogen
molecule into account, are still only possible if all symmetries of
the scattering problem are utilized.

The classical trajectory calculations are performed on 
{\em exactly the same} PES as the quantum dynamical calculations.
The equations of motions are numerically integrated with the
Bulirsch-Stoer method with a variable time-step \cite{NumRec}. 
The energy conservation is typically fulfilled to 0.1~meV.
The sticking probability is determined by averaging over
a sufficient number of trajectories. The exact number of trajectories 
depends on the specific initial conditions and ranges between 1,815 and
18,330.

As far as the CPU time requirement is concerned, it is a wide-spread
believe that classical methods are much less time-consuming than
quantum ones. This is certainly true if one compares the computational
cost of one trajectory to a quantum calculations. However, if one is
interested in averaged quantities like sticking probablities, then 
in classical molecular dynamics calculations one
has to average over many trajectories corresponding to different
initial conditions. Quantum mechanics does this averaging automatically.
A plane wave describing the incident beam hits the surface everywhere, 
and a $j=0$ rotational state contains all molecular orientations.
Thus for the results presented here the quantum method is even
more time-efficient than the classical calculations, in particular if
one considers the fact, that in a coupled-channel method the
sticking and scattering probabilites of all open channels are
determined simultaneously.

\section{Results and Discussion}

\begin{figure}[t]
\unitlength1cm
\begin{center}
   \begin{picture}(10,8.5)
   \includegraphics{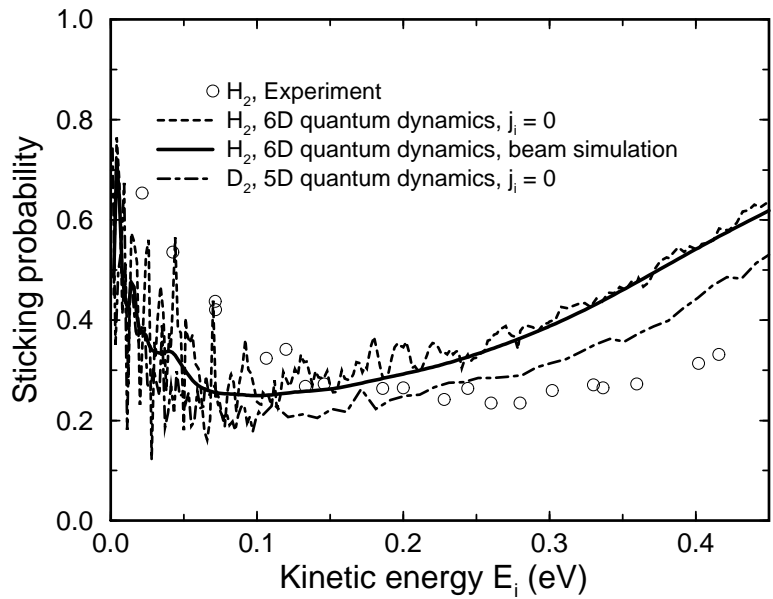}
   \end{picture}
\end{center}
   \caption{Sticking probability versus kinetic energy for
a hydrogen beam under normal incidence on a Pd(100) surface.
Experiment (H$_2$): circles (from ref.~\protect{\cite{Ren89}}); theory:
six-dimensional results for H$_2$ molecules initially in the rotational 
and vibrational ground state (dashed line)
and with an initial rotational and energy distribution 
adequate for molecular beam experiments (solid line), and 
vibrationally adiabatic five-dimensional
results for D$_2$ molecules initially in the rotational ground state 
(dash-dotted line).}
\label{stickiso}
\end{figure}

Figure \ref{stickiso} presents six-dimensional quantum dynamical
calculations of the sticking probability  as a function of the 
kinetic energy of a H$_2$ beam under normal incidence on a Pd(100) surface 
and five-dimensional calculations for D$_2$. In addition, the results 
of a H$_2$ molecular beam experiment are shown \cite{Ren89}.
Quantum mechanically determined sticking probabilities 
for hydrogen at surfaces with an attractive well exhibit an oscillatory
structure as a function of the incident energy~\cite{Gro95,Kay95,Dar90,Gro95b},
reflecting the opening of new scattering channels and 
resonances \cite{Dar90,Gro95b}. These structures are known for a long time
in He and H$_2$ scattering \cite{stern} and also in LEED \cite{LEED}. For 
H$_2$/Pd(100), however, measuring these oscillations is a very demanding
task. They are very sensitive to surface imperfections like adatoms
or steps \cite{Gro96c} and therefore hard to detect \cite{Ret96}. 
Since we do not focus on these oscillations here, for the solid line in
Fig.~\ref{stickiso} we have assumed a velocity spread of the incoming
beam typical for the experiment \cite{Ren89} so that the oscillations
are smoothed out.

The initial decrease of the sticking probability found in the
experiment is well-produced in the quantum dynamical calculations.
The high sticking probability
at low kinetic energies is caused by a steering effect: Slow
molecules can very efficiently be steered to non-activated pathways
towards dissociative adsorption by the attractive forces of the
potential. This mechanism becomes less effective at higher 
kinetic energies where the molecules are too fast to be focused
into favourable configurations towards dissociative adsorption.
This causes the initial decrease of the sticking probability. If the
kinetic energy is further increased, the molecules will eventually
have enough energy to directly traverse the barrier region leading
to the final rise in the sticking probability.

\begin{figure}[tb]
\unitlength1cm
\begin{center}
   \begin{picture}(10,8.5)
      \includegraphics{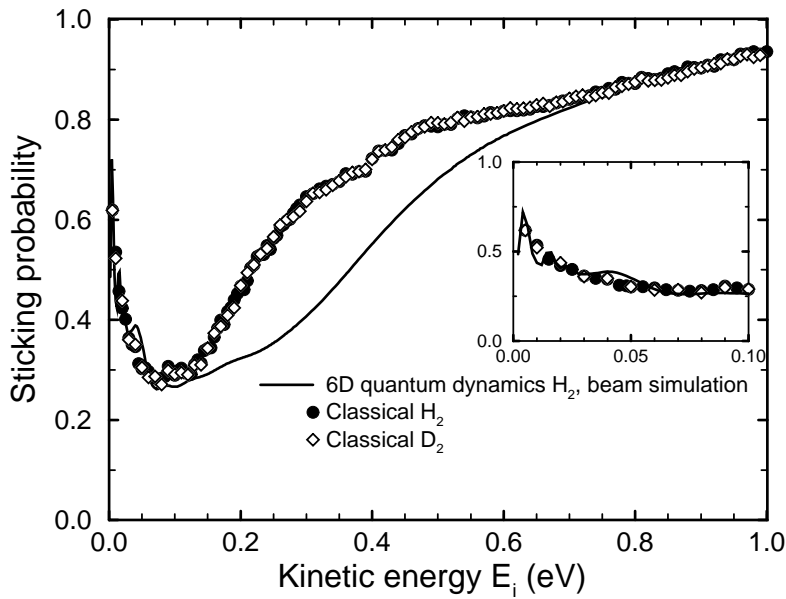}
   \end{picture}

\end{center}
   \caption{Probability for dissociative adsorption
versus kinetic translational energy for a H$_2$ beam under normal
incidence on a clean Pd(100) surface for non-rotating molecules.
The solid line corresponds to the six-dimensional quantum dynamical 
results assuming an energy spread typical for beam experiments. 
The molecular dynamics results (H$_2$: circles, D$_2$: diamonds)
have been obtained purely classical, i.e. without including 
zero-point energies in the initial conditions.
The inset shows an enlargement of the results at low energies.}
\label{stickclass}
\end{figure}

In Fig.~\ref{stickclass} we compare the averaged quantum mechanical
sticking probability for H$_2$ with the results of classical trajectory
calculations for H$_2$ and D$_2$. 
The inset shows an enlargement of the results at low energies.
The molecular dynamics calculations have been performed {\em without}
any zero-point energies in the initial conditions. This allows us
to truely differentiate between classical and quantum effects in
the dynamics. First of all, the classical results do not show
any oscillatory structure revealing that the oscillations are purely
due to quantum mechanics. Furthermore, at low energies the classical
results fall almost exactly upon the averaged quantum results.
This shows that steering is a general concept and not restricted
to quantum or classical dynamics. We attribute the difference between the 
classical and quantum results at the medium energy range to the
influence of zero-point effects in the multi-dimensional interaction
potential \cite{Gro96d}. 
At very high energies, where zero-point efects play no role,
quantum and classical results again agree.

\begin{figure}[ht]
\unitlength1cm
\begin{center}
   \begin{picture}(10,6.5)
      \includegraphics{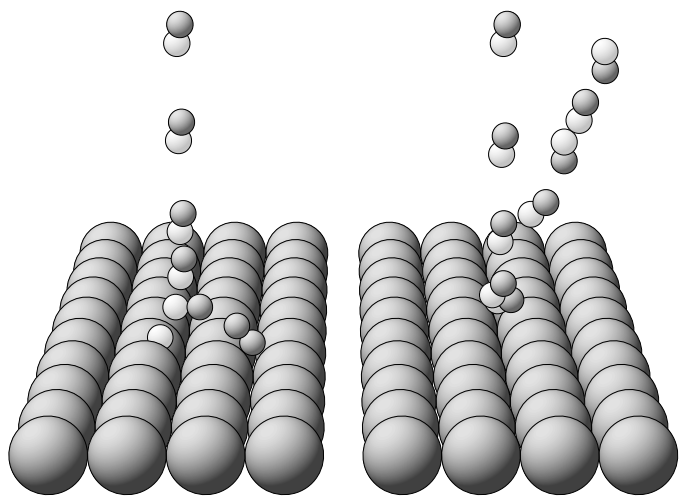}
   \end{picture}

\end{center}
   \caption{Snapshots of classical trajectories of hydrogen molecules 
impinging on a  Pd(100) surface. The initial conditions are chosen in 
such a way that the trajectories are restricted to the $xz$-plane.
Left trajectory: initial kinetic energy $E_i = 0.01$~eV. 
Right trajectory: same initial conditions as in the left trajectory
except that the molecule has a higher kinetic energy of 0.12 eV. }

\label{traj2run}
\end{figure}

Since the averaged quantum and the classical sticking probability
agree in the low-energy regime, we can use classical trajectories
to describe the steering mechanism. This is done in 
Fig.~\ref{traj2run}, where snapshots of two typical trajectories
are shown.  The initial conditions are chosen in 
such a way that the trajectories are restricted to the $xz$-plane.

The left trajectory illustrates the steering effect \cite{Gro95,Kay95}. 
The initial kinetic energy is $E_i = 0.01$~eV. 
Initially the molecular axis is almost perpendicular to the
surface. In such a configuration the molecule cannot dissociate
at the surface. But the molecule is so slow that the attractive
forces can reorient the molecule so that it can follow a non-activated
path towards dissociative adsorption. 

In the case of the right trajectory, the initial conditions are the same
as in the left one, except  that the molecule has a higher kinetic 
energy of 0.12~eV. Due to the anisotropy of the PES the molecule also
starts to rotate to a configuration parallel to the surface. However,
now the molecule is so fast that it hits the repulsive wall of the potential
before it is in a favorable configuration to dissociative adsorption.
It is then scattered back into the gas-phase rotationally excited.

Finally we like to discuss the isotope effects in the dissociation
of hydrogen on Pd(100).
Fig.~\ref{stickclass} shows that in clasical dynamics there is practically
no isotope effect between H$_2$ and D$_2$ in the sticking probability.
As far as the low-energy regime is concerned, this seems surprising at
a first glance, since D$_2$ is more inert than H$_2$ due to its higher mass.
However, one has to keep in mind that at the same kinetic energy D$_2$
is slower than H$_2$, so that there is more time for the steering forces 
to redirect the D$_2$ molecule. This compensating effect has also been
found theoretically for the dissociation of hydrogen at W(100) \cite{Kay95}.

In Fig.~\ref{stickiso} we have also plotted the sticking probability
of D$_2$ according to five-dimensional quantum dynamical calculations.
Due to its higher mass the energy spacing between the quantum levels
is smaller for D$_2$ than for H$_2$. Therefore much more eigenfunctions
in the expansion of the wavefunction have to be taken into account in
the coupled-channel calculations for D$_2$ than for H$_2$. This makes
a six-dimensional quantum treatment of D$_2$ not feasible at the moment.
However, we have already shown that the results of five-dimensional
vibrationally adiabatic quantum calculations, where the molecules
are not allowed to make vibrational transitions, are very close to
the full six-dimensional results for the dissociation
of hydrogen on Pd(100)~\cite{Gro96b}.  
Hence it is reasonable to compare five-dimensional results for D$_2$
with six-dimensional results for H$_2$.
The quantum dynamical sticking probabilities of D$_2$ are 
slightly smaller than those of H$_2$. Since no such isotope effect is observed
in the classical calculations (Fig.~\ref{stickclass}), this small difference
has to be a quantum mechanical effect. 
We attribute it to the larger vibrational zero-point energy of H$_2$ which
can effectively be used to traverse the barrier region \cite{Gro96b,Gro96d}.

\section{Conclusions}

In conclusion, we reported a six-dimensional quantum and
classical dynamical study of dissociative adsorption of
H$_2$/Pd\,(100). We have shown that the initial decrease of
the sticking probability is due to a steering mechanism
which is operative both in quantum and classical dynamics.
The adsorption dynamics shows only a small isotope effect which is purely
due to the quantum nature of hydrogen.  Our results establish
the importance of a high-dimensional dynamical treatment in order
to understand reactions at surfaces.


\begin{thebibliography}{99}


\small\normalsize

\bibitem{Ren94} K.D. Rendulic and A. Winkler, Surf. Sci. {\bf 299/300}, 261
(1994). 
\bibitem{Hol94} S. Holloway, Surf. Sci. {\bf 299/300}, 656 (1994).
\bibitem{Dar95} G.R. Darling and S. Holloway, Rep. Prog. Phys. {\bf 58}, 1595 
(1995).
\bibitem{Gro96SS} A. Gross, Surf. Sci. {\bf 363}, 1 (1996).


\bibitem{Ham94} B. Hammer, M. Scheffler, K.W. Jacobsen, and J.K. N{\o}rskov,
Phys. Rev. Lett. {\bf 73}, 1400 (1994).
\bibitem{Whi94} J.A. White, D.M. Bird, M.C. Payne, and I. Stich,
Phys. Rev. Lett. {\bf 73}, 1404 (1994).
\bibitem{Wil95} S. Wilke and M. Scheffler, Surf. Sci. {\bf 329}, L605 (1995);
                Phys. Rev. B {\bf 53}, 4926 (1996).
\bibitem{Wil96} S. Wilke and M. Scheffler, Phys. Rev. Lett. {\bf 76}, 
3380 (1996)
\bibitem{Wie96} G. Wiesenekker, G.J. Kroes, and E.J. Baerends, J. Chem. Phys.
{\bf 104}, 7344 (1996).


\bibitem{Gro95} A. Gross, S. Wilke, and M. Scheffler, Phys. Rev. Lett. 
{\bf 75}, 2718 (1995).


\bibitem{Ren89} K. D. Rendulic, G. Anger, and A. Winkler, Surf. Sci. {\bf 208},
404 (1989). 
\bibitem{Res94} Ch. Resch, H. F. Berger, K. D. Rendulic, and E. Bertel, 
Surf. Sci. {\bf 316}, L1105 (1994).
\bibitem{Ber92} H. F. Berger, Ch. Resch, E. Gr\"osslinger, G. Eilmsteiner, A.
Winkler, and K. D. Rendulic, Surf. Sci. {\bf 275}, L627 (1992).
\bibitem{But94} D. A. Butler, B. E. Hayden, and J. D. Jones, Chem. Phys. Lett.
{\bf 217}, 423 (1994).
\bibitem{Aln89} P. Alnot, A. Cassuto, and D. A. King, Surf. Sci. {\bf 215}, 29
(1989).
\bibitem{But95} D. A. Butler and B. E. Hayden, Chem. Phys. Lett.
{\bf 232}, 542 (1995).
\bibitem{Dix94} St. J. Dixon-Warren, A. T. Pasteur, and D. A. King, 
Surf. Rev. and Lett. {\bf 1}, 593 (1994).


\bibitem{Kay95} M. Kay, G.R. Darling, S. Holloway, J.A. White, and D.M. Bird,
Chem. Phys. Lett. {\bf 245}, 311 (1995).

\bibitem{Gro96} A. Gross, S. Wilke, and M. Scheffler, 
Surf. Sci. {\bf 357/358}, 614 (1996). 


\bibitem{Gro96b} A. Gross and M. Scheffler, Chem. Phys. Lett. {\bf 256},
417 (1996).




\bibitem{Per92}J. P. Perdew J. A. Chevary, S. H. Vosko, K. A. Jackson,
M. R. Pederson, D. J. Singh, and C. Fiolhias, 
Phys.~Rev.~B {\bf 46}, 6671 (1992).
\bibitem{Bla93} P. Blaha, K. Schwarz, and R. Augustyn, WIEN93, Technical
University of Vienna 1993.
\bibitem{Koh95} B. Kohler, S. Wilke, M. Scheffler, R. Kouba, and 
C. Ambrosch-Draxl, Comput. Phys. Commun. {\bf 94}, 31 (1996).
\bibitem{Bre93} W. Brenig, T. Brunner, A. Gross, and R. Russ, Z.~Phys.~B 
{\bf 93}, 91 (1993).

\bibitem{NumRec} W.H. Press, B.P. Flannery, S.A. Teukolsky, and 
W.T. Vetterling, {\em Numerical Recipes}, Cambridge University Press,
Cambridge, 1989.



\bibitem{Dar90} G.R. Darling and S. Holloway, J. Chem. Phys. {\bf 93}, 9145 
(1990). 
\bibitem{Gro95b} A. Gross, J. Chem. Phys. {\bf 102}, 5045 (1995).

\bibitem{stern}  R. Frisch and O. Stern, Z. Phys. {\bf 84}, 430 (1933).
\bibitem{LEED}  J.B. Pendry, {\it Low energy electron diffraction}, Academic
Press, London (1974), p. 112.
\bibitem{Gro96c} A. Gross and M. Scheffler, Phys. Rev. Lett. {\bf 77}, 
405 (1996).


\bibitem{Ret96} C.T. Rettner and D.J. Auerbach, Chem. Phys. Lett. {\bf 253},
236 (1996).

\bibitem{Gro96d} A. Gross and M. Scheffler, to be published.









\end{thebibliography}
\end{document}